\documentclass[12pt]{article}
\usepackage{url}
\def\@preprint{\relax}
\newcommand\preprint[1]{\long\gdef\@preprint{#1}}
\catcode`\@=11

\@addtoreset{equation}{section}
\def\theequation{\arabic{equation}}
\def\theequation{\thesection\arabic{equation}}

\newcommand{\appendixA}{\setcounter{equation}{0}
\def\theequation{\rm{A}.\arabic{equation}}\section*}
\newcommand{\appendixB}{\setcounter{equation}{0}
\def\theequation{\rm{B}.\arabic{equation}}\section*}

\catcode`\@=11

\def\section{\@startsection{section}{1}{\z@}{3.5ex plus 1ex minus
   .2ex}{2.3ex plus .2ex}{\large\bf}}
\def\thesection{\arabic{section}.}

\def\ps@headings{\def\@oddfoot{}\def\@evenfoot{}
\def\@oddhead{\hbox{}\hfill
        \makebox[.5\textwidth]{\raggedright\ignorespaces --\thepage{}--
        \hfill }}
\def\@evenhead{\@oddhead}
\def\subsectionmark##1{\markboth{##1}{}} }
\def\be{\begin{equation}}
\def\ee{\end{equation}}
\def\ba{\begin{eqnarray}}
\def\ea{\end{eqnarray}}
\def\bs{\begin{subequations}}
\def\es{\end{subequations}}

\newcommand\fverb{\setbox\pippobox=\hbox\bgroup\verb}
\newcommand\fverbdo{\egroup\medskip\noindent%
                        \fbox{\unhbox\pippobox}\ }
\newcommand\fverbit{\egroup\item[\fbox{\unhbox\pippobox}]}
\newbox\pippobox

\def\co{{\cal O}}
\def\e{\epsilon}

\def\sp{\;\;\;,\;\;\;}
\def\spa{\;,\;}

\begin{document}
\thispagestyle{empty} \rightline{\normalsize\sf hep-th/0204153}
\rightline{\normalsize CERN-TH/2002-083} \rightline{\normalsize LPTENS-02-21}
\vskip .4truecm \centerline{\Large\bf Anomalous U(1)s in type I string vacua}
\vskip .6truecm
\begin{center}
{{\bf Ignatios Antoniadis}\footnote{On leave of absence from \it
CPHT, UMR du CNRS 7644, Ecole Polytechnique, F-91128 Palaiseau}\\
Theory Division, CERN\\
CH 1211 Geneva 23, Switzerland\\
{\tt E-mail: antoniad@mail.cern.ch}}
\\
{{\bf Elias Kiritsis}\\ Laboratoire de Physique Th\'eorique
      de l'Ecole Normale Sup\'erieure\\
      24 rue Lhomond,
      Paris, Cedex 05, F-75231, FRANCE\\and\\
Department of Physics, University of Crete, and FO.R.T.H.\\
71003 Heraklion, Greece\\
{\tt E-mail: kiritsis@physics.uoc.gr}}\\
{{\bf John Rizos}\\
 Department of Physics, University of Ioannina,\\ 45110 Ioannina,
Greece\\
{\tt E-mail: irizos@cc.uoi.gr}}
\end{center}
\centerline{\bf\small ABSTRACT}
\vskip .4truecm
{We perform a systematic string computation of the masses of
anomalous $U(1)$ gauge bosons in four-dimensional orientifold
vacua, and we study their localization properties in the internal
(compactified) space. We find that $N=1$ supersymmetric sectors
yield four-dimensional contributions, localized in the whole
six-dimensional internal space, while $N=2$ sectors give
contributions localized in four internal dimensions. As a result,
the $U(1)$ gauge fields can be much lighter than the string scale,
so that when the latter is at the TeV, they can mediate new
non-universal repulsive forces at submillimeter distances much
stronger than gravity. We also point out that even $U(1)$s which
are free of four-dimensional anomalies may acquire non-zero masses
as a consequence of six-dimensional anomalies.}
\hfill\break
\vfill\eject

\section{Introduction}

Anomalous $U(1)$ gauge symmetries appear generically in string
vacua. The massless charge spectrum is anomalous in the sense that the
traditional triangle (or polygon in dimensions higher than four)
diagrams are non-zero. However, the anomaly is cancelled via a
generalization of the Green-Schwarz mechanism \cite{Green:sg, Dine:xk}.
In four dimensions, a scalar axion (zero-form, or its dual two-form)
is responsible for the anomaly cancellation. In six dimensions, both
zero-forms (or their duals four-forms) and two-forms can participate in
anomaly cancellation \cite{Sagnotti:1992qw}. However, only zero-forms (or
their duals) can give mass to the $U(1)$ gauge boson and break the gauge
symmetry. Thus, a necessary consequence in four dimensions is that the
(quasi)anomalous gauge symmetry is broken. Moreover, in the presence of
supersymmetry (at least), an anomalous $U(1)$ is accompanied by a D-term
potential that involves the charged scalars, shifted by a term proportional
to the CP-even partner of the respective axion \cite{Dine:xk}.

In perturbative heterotic vacua, at most one
anomalous $U(1)$ can appear in four-dimensio\-nal $N=1$
compactifications. The relevant axion is the four-dimensional dual of the
Neveu-Schwarz (NS) two-form, which was shown to develop the appropriate
couplings and transformation properties, needed to cancel all relevant
anomalies \cite{Dine:xk}. Moreover, the scalar modulus appearing in the
D-term potential is the dilaton, and for non-trivial vacua, vanishing of
the D-term implies generically that charged scalars get a non-trivial
vacuum expectation value (VEV) breaking the associated global $U(1)$
symmetry.

The situation is richer and more
interesting in perturbative orientifold vacua. Here, there are in
general several anomalous $U(1)$s and the cancellation of
anomalies is achieved via the coupling of twisted Ramond-Ramond (RR) axions
\cite{Ibanez:1999qp}. The D-term potentials involve the twisted
NS-NS moduli. However, at the orientifold point their expectation
values vanish, and this allows to have a spontaneously broken
gauge symmetry with the global $U(1)$ unbroken in
perturbation theory \cite{Poppitz:1998dj}. The global symmetry may be broken
non-perturbatively due to instanton effects, which however are
small at weak coupling.

Orientifold vacua are prime candidates for realizing the Standard
Model as a brane-world, in the context of perturbative string
theory with low string scale and large internal dimensions
\cite{Antoniadis:1998ig} (for earlier attempts see
\cite{Antoniadis:1990ew}). As pointed out in
\cite{Antoniadis:2000en}, any minimal realization of the Standard
Model in this context contains at least 2 anomalous $U(1)$s that
are expected to obtain a mass. Abelian gauge symmetries have been
also used on the world-brane or in the bulk, in order to impose
approximate global symmetries, such as baryon number or
Peccei--Quinn symmetries \cite{Arkani-Hamed:1998nn,Ibanez:1999it}.
It is therefore important to compute their masses and study their
localization properties in the internal compact space.

It turns out that the mass of anomalous $U(1)$s in orientifold
vacua can be unambiguously calculated by a direct one-loop string
computation (although a disk calculation may also give the mass modulo
normalization ambiguities). In this work, we perform such a computation
and we derive a formula for the mass matrix of
$U(1)$ gauge bosons. We also study some explicit examples of
$Z_N$ and $Z_N\times Z_M$ orientifold vacua.

We find the following general features:
\begin{enumerate}
\item The gauge boson masses are given by an ultraviolet contact term
of the one-loop annulus diagram with the gauge bosons inserted one
at each boundary. There are no contributions from the annulus
with insertions on the same boundary or from the M\"obius strip since
such contact terms are absent by tadpole cancellation. By
open-closed string duality, the $U(1)$ mass-terms are also given by some
appropriate infrared (IR) closed string channel tadpoles.

\item The mass-terms of $U(1)$ gauge bosons obtain volume independent
corrections from $N=1$ supersymmetric sectors, while $N=2$ sectors give
contributions dependent on the moduli of the corresponding fixed torus.
Moreover, they are BPS saturated (given by the
supertrace of the square of the four-dimensional helicity).
Thus, mass-terms of $N=1$ sectors are localized in all six internal
dimensions, while those of $N=2$ sectors are six-dimensional, localized in
four internal dimensions.

\item $U(1)$s that are free of four-dimensional anomalies can still be
massive, if upon decompactification they suffer from six-dimensional
anomalies\footnote{Similar observations were made independently in
\cite{Scrucca:2002is}.}. This is expected since Kaluza--Klein (KK) states can
contribute to (higher-dimensional) anomalies once there is a corner of
moduli
space where they become massless (decompactification limit). Uncancelled
anomalies in six dimensions depend both on the localization of gauge fields
(D-branes) and the localization of axions (coming from the bulk).
Potentially
anomalous sectors involve a six-dimensional coupling of a gauge field to an
axion that extend in the same six dimensions.

\end{enumerate}

As we already mentioned, the masses of $U(1)$s arise through Green-Schwarz
couplings involving RR axions. Moreover, at the orientifold point, the
associated global symmetries remain unbroken to all orders in perturbation
theory. Using the localization properties we described above, one can
provide
explicit realizations of all possible arrangements for the abelian gauge
bosons
$(A)$ and their corresponding axions $(a)$: \be (A,a)={\rm (brane, brane)\,
,\
(bulk, brane)\, ,\ (brane, bulk)\, ,\ (bulk, bulk)}\, . \label{cases} \ee
$N=1$
sectors realize the first two possibilities, while $N=2$ sectors realize the
last two. Note that the axions can propagate at most in two internal
dimensions, while $U(1)$ gauge bosons may propagate everywhere. It follows
that
the $U(1)$ mass $M_A$ in these four cases is proportional to: \be M_A\sim
{\cal
O}(1)\ ,\quad 1/\sqrt{V_A}\ ,\quad 1/\sqrt{V_a}\ ,\quad \sqrt{V_a/V_A}
\label{MA} \ee in string units, where $V_A$ and $V_a$ stand for the internal
volumes corresponding to the propagation of the $U(1)$ and the axion fields,
respectively ($V_a$ is two-dimensional).

As a result, $M_A$ can vary from the string scale $M_s$, up to
much lower values that can attain $M_s^2/M_{\rm Planck}$, in the
two middle cases of (\ref{cases}), if the respective volume in
eq.(\ref{MA}) coincides with the total volume of the bulk. The
gauge field exchange can then induce new (repulsive) forces at
sub-millimeter distances (of the order of a few microns for $M_s$
a few TeV). The third case, where the gauge field lives on the
brane, is however experimentally excluded, since the corresponding
gauge coupling $g_A$ is of order unity. In the second case, the
gauge field lives in the bulk and the four-dimensional $U(1)$
gauge coupling is infinitesimally small, $g_A\sim M_s/M_{\rm
Planck}\simeq 10^{-16}$. However, this value is still bigger that
the gravitational coupling $\sim E/M_P$ for typical energies $E$
of the order of the proton mass, and the strength of the new force
would be $10^6-10^8$ stronger than gravity. This an interesting
region which will be soon explored in micro-gravity experiments
\cite{exp}. Notice that the supernova constraints can exclude only
the case where there are less than four large extra dimensions in
the bulk, felt by the gauge field \cite{Arkani-Hamed:1998nn}.
Finally, in the (bulk, bulk) case, the masses of all KK modes are
shifted by a large amount according to eq.(\ref{MA}) and the
resulting force becomes effective at much smaller distances.

Of course, in all cases of (\ref{cases}), the $U(1)$ gauge bosons can be
produced in particle accelerators at high energies leading to interesting
signatures. Note that their masses are always lower than the string scale
because of the (string) one-loop factor suppression. Moreover, their
effective coupling is of order unity, if one takes into account the number
of KK excitations that are produced at high energies.

The paper is organized as follows. In Section 2, we describe the effective
action involving the anomalous $U(1)$ symmetries. In Section 3, we present
the one-loop string computation and we give the general results for the
contributions of $N=1$ and $N=2$ supersymmetric orbifold sectors. In Section
4, we study specific orientifold examples based on $Z_3$, $Z_7$, $Z_2\times
Z_3=Z_6'$, $Z_6$ and $Z_3\times Z_6$ orbifolds. Finally, Section 5 contains
concluding remarks and a discussion of non supersymmetric models.

\section{The effective action}
\setcounter{equation}{0}

In four dimensions, there are two on-shell equivalent (dual) ways
of describing the fields responsible for cancelling anomalies: as
pseudo-scalars or as two-index antisymmetric tensors. However,
off-shell, the two descriptions are a priori different at the
one-loop level.

Let us first consider the case of a pseudoscalar axion. The relevant part
of the four-dimensional effective action (in the string frame) is:
\be
S=\int d^4 x\left[-{1\over 4g_A^2}F_A^2-{1\over
2}(da+M A)^2 +{a\over M}\sum_I k_I F_I\wedge F_I
\right]\, ,
\label{actiona}
\ee
where $F_A$ is the field strength of the anomalous $U(1)_A$, $g_A$ is the
corresponding gauge coupling, and $k_I$ are the various mixed anomalies.
Anomaly cancellation implies that $a$ is shifted under $U(1)_A$ gauge
transformation: $\delta A=d\Lambda$, $\delta a=-M\Lambda$, so that
the action (\ref{actiona}) changes by exactly the
amount necessary to cancel the phase of the chiral fermion determinant. It
follows that in the unitary gauge $a$ vanishes and one is left over with a
massive $U(1)_A$ with mass $M_A=g_A M$.
Note that in the case where the gauge
symmetry is not anomalous in four dimensions but the gauge field becomes
massive due to a six-dimensional anomaly, all $k_I$ vanish but still $M\ne
0$ and $a$ transforms under gauge transformation.

In the type I string context, where the axion $a$ comes from the RR closed
string sector \cite{Ibanez:1999qp}, the first and third terms of the above
effective action appear at the level of the disk, while the second term is
expanded into contributions corresponding to different orders of string
perturbation theory; $(da)^2$ is a tree-level (sphere) term, the
cross-product
$Ada$ appears at the disk level, while the mass-term $A^2$ is a one-loop
contribution. Indeed, for this counting, the gauge kinetic terms have a
dilaton
factor $e^{-\phi}$ since $g_A^2$ is proportional to the string coupling
$g_s\equiv e^\phi$, while a one-loop term is dilaton independent. On the
other
hand, every power of the RR field $a$ absorbs a dilaton factor $e^{-\phi}$
which makes both the last two terms in (\ref{actiona}) dilaton independent.

In a supersymmetric theory (at least), the above effective action is
accompanied by a D-term potential:
\be
V=\int d^4 x {1\over g_A^2}{\rm D}^2\quad ;\qquad
{\rm D}=Mm+\sum_iq_i|\Phi_i|^2\, ,
\label{pot}
\ee
where $m$ is the twisted NS-NS blowing-up modulus that belongs in the same
chiral multiplet with the RR axion $a$, while $\Phi_i$ denote the various
open string charged scalars with $U(1)_A$ charges $q_i$. At the orientifold
point, $m$ vanishes and the global $U(1)_A$ symmetry remains unbroken
despite the fact that the gauge field $A$ becomes massive for $M\ne 0$
\cite{Poppitz:1998dj}. However, going away from the orientifold point when
$m\ne 0$, vanishing of the D-term implies that some charged scalars should
acquire non-zero VEVs, breaking the global $U(1)_A$ symmetry.

The above phenomenon can also be described in terms of an antisymmetric
tensor
$B_{\mu\nu}$. The corresponding effective action can be easily obtained from
(\ref{actiona}) by performing a standard Poincar\'e duality which exchanges
equations of motion with Bianchi identities \cite{Klein:2000im}: \be S=\int
d^4
x\left[-{1\over 4g_A^2}F_A^2-{1\over 12} (dB+{k_I\over M}\Omega_I)^2 +
(MdB+k_I
\Omega_I)\wedge A \right]\, , \label{actionB} \ee where $\Omega_I$ are the
various gauge Chern--Simons terms. Anomaly cancellation implies that
$B_{\mu\nu}$ is shifted under gauge transformations: $\delta
B=\sum_I\Lambda_I
k_I F_I/M$, while the variation of the last term under $U(1)_A$ gauge
transformation, cancels the anomalous contribution of the chiral fermion
determinant.

The counting of the order of appearance of the various terms in type I
string perturbation theory is similar as in the dual action
(\ref{actiona}). However, notice that in this representation there is no
explicit mass-term for the $U(1)_A$ gauge field in the effective action.
The reason is that the mass is now generated by a reducible diagram at the
one-loop level. Inspection of (\ref{actionB}) shows that a mixing between
$B_{\mu\nu}$ and $A_{\rho}$ arises at the level of the disk, corresponding
to the vertex ${M\over 2}
\epsilon^{\mu\nu\rho\sigma}\epsilon_{\mu\nu}\epsilon_{\rho}p_{\sigma}$,
where $p$ is the external momentum and $\epsilon$'s denote the polarization
tensors. This generates a reducible contribution to the two-point function
of the gauge field, given by the square of this vertex times the
$B_{\mu\nu}$ propagator:
\be
{M^2\over 2}\left[p^2~\epsilon^2-(p\cdot \e)^2\right]/p^2=
{M^2\over 2}\epsilon^2\, ,
\ee
where we have used the on-shell gauge-invariance
condition $p\cdot \e=0$. Thus, a mass $M$ for the anomalous gauge boson
is generated by such a reducible one-loop diagram and no explicit
mass-term is present in the effective lagrangian.

Note that in the axion representation (\ref{actiona}), the axion-gauge
boson mixing at the disk-level generates a
vertex proportional to $p\cdot \e$ which vanishes on-shell and
does not generate a reducible contribution for the gauge boson mass.
Thus, an explicit one-loop mass-term must be introduced in the
effective lagrangian, consistently with the expression (\ref{actiona}).

\section{The calculation of the mass in orientifold models}
\setcounter{equation}{0}

The two possible diagrams that can contribute to terms quadratic in the
gauge boson at the one-loop level are the annulus and the M\" obius strip.
Of those, only the annulus with the gauge field vertex operators inserted
at the two opposite ends has the appropriate structure to contribute to
the mass-term. Indeed, vertex operators inserted at the same boundary will
be proportional to $Tr[\gamma_{k}\lambda^a\lambda^b]$, where $\gamma_k$ is
the representation of the orientifold group element in the $k$-th orbifold
sector acting on the Chan--Paton (CP) matrices $\lambda ^a$.
On the other hand, for gauge fields inserted on opposite boundaries, the
amplitude will be proportional
to $Tr[\gamma_{k}\lambda^a]Tr[\gamma_{k}\lambda^b]$ and it is this form of
traces that determines the anomalous $U(1)$s \cite{Ibanez:1999qp}. The
potential ultraviolet (UV) divergences that come from vertex operators
inserted on the same boundary (both in cylinder and M\"obius strip) cancel
by tadpole cancellation \cite{Antoniadis:1999ge}.

Obviously, we must concentrate on the {\em CP}-even part of the amplitude
which receives contributions only from even spin structures. This implies
that we need the gauge boson vertex operators in the zero-ghost picture:
\be
V^a=\lambda^a\epsilon_{\mu}(\partial X^{\mu}+
i(p\cdot \psi)\psi^{\mu})e^{ip\cdot X}\, ,
\ee
where $\lambda$ is the Chan--Paton matrix and $\epsilon^{\mu}$ is the
polarization vector.

The world-sheet annulus is parameterized by $\tau=it/2$, where
$\tau=\tau_1+i\tau_2$ is the usual complex modular parameter of the torus,
and corresponds to the rectangle $[0,t/2]\otimes[0,1/2]$.
The 2-point amplitude is then given by \cite{Bain:2000fb}:
\be
{\cal A}=-{1\over 4|G|}\int [d\tau][dz]
\int {d^4p\over (2\pi)^4}\sum_{k}\langle
V(\epsilon_1,p_1,z)V(\epsilon_2,p_2,z_0)\rangle_k\, ,
\ee
where the sum is over orientifold sectors, $|G|$ is the order of the
orientifold group, and we fixed one of the positions of the vertex
operators at $z_0=1/2$ using the translational symmetry of the
annulus. The other vertex operator is located on the opposite boundary:
$z=i\nu$ with $\nu\in [0,t/2]$. For notational simplicity, we set the Regge
slope $\alpha'=1/2$ so that the Virasoro Hamiltonian operator
$L_0=(p^2+M^2)/2$.

Performing the contractions, we obtain
$$
{\cal A}=-{1\over 2|G|}\int [d\tau][dz]\int {d^4p\over (2\pi)^4}\sum_{k}
[(\epsilon_1\cdot \epsilon_2)(p_1\cdot p_2)-
(\epsilon_1\cdot p_2)(\epsilon_2\cdot p_1)] (Tr[\gamma_k\lambda])^2
$$
\be
\times
e^{-p_1\cdot p_2\langle X(z)X(z_0)\rangle}\left[
\langle\psi(z)\psi(z_0)\rangle^2-\langle X(z)\partial
X(z_0)\rangle^2\right]\, .
\ee
It appears that the amplitude is $\co(p^2)$ and thus provides a
correction only to the anomalous gauge boson coupling.
We will see however, that after integration over the position $z$ and the
annulus modulus $t$, a term proportional to $1/ p_1\cdot p_2$ appears from
the ultraviolet (UV) region (as a result of the quadratic UV
divergence in the presence of anomalous $U(1)$s) that will provide the
mass-term.

Strictly speaking, the amplitude above is zero on-shell if we enforce the
physical state conditions $\epsilon\cdot p=p^2=0$ and momentum conservation
$p_1+p_2=0$. There is however a consistent off-shell extension, without
imposing momentum conservation, that has given consistent results in other
cases (see \cite{Kiritsis:1998hj} for a discussion) and we adopt it here.
We will thus impose momentum conservation only at the end of the
calculation. We now define for convenience the reduced amplitude ${\cal
A}_k$ by amputating the kinematical factors\footnote{We consider in
general insertions of different gauge fields on the different boundaries.
The gauge fields can belong to different types of branes.}
\be
{\cal A}^{ab}=\int {d^4p\over (2\pi)^4}[(\epsilon_1\cdot
\epsilon_2)(p_1\cdot p_2)-
(\epsilon_1\cdot p_2)(\epsilon_2\cdot p_1)]\sum_k
~Tr[\gamma_k\lambda^a]Tr[\gamma_k\lambda^b]~~{\cal A}^{ab}_{k}
\ee
with
\be
{\cal A}^{ab}_k=-{1\over 2|G|}\int [d\tau][dz]e^{-\delta
\langle X(z)X(z_0)\rangle}\left[ \langle\psi(z)\psi(z_0)\rangle^2-
\langle X(z)\partial X(z_0)\rangle^2\right]Z^{ab}_{k}\, ,
\label{ampldef}
\ee
where $Z^{ab}_{k}$ is the annulus partition function in the $k$-th orbifold
sector, and we have set $\delta\equiv p_1\cdot p_2$.
The dependence of ${\cal A}^{ab}_k$ on the two gauge indices $a,b$ is mild.
It depends only on the type of brane the gauge fields come from. For
instance, in standard supersymmetric $Z_N$ orbifolds, there are three
different cases corresponding to 99, 55 and 95 D-brane combinations.

We will need here the bosonic and fermionic propagators on the annulus.
They can be obtained from those of the torus:
\be
\langle X(e^{2\pi i\nu_1})X(e^{2\pi i\nu_2})\rangle=-{1\over 4}
\log\left|{\vartheta_1(\nu_1-\nu_2|\tau)\over
\vartheta'_1(0|\tau)}\right|^2+{\pi{\rm Im}^2(\nu_1-\nu_2)\over 2\tau_2}
\ee
\be
\langle \psi(e^{2\pi i\nu_1})\psi(e^{2\pi i\nu_2})
\rangle\left(^{\alpha}_{\beta}\right)
={i\over 2}{\vartheta\left(^{\alpha}_{\beta}\right)(\nu_1-\nu_2|\tau)
\vartheta'_1(0|\tau)\over
\vartheta\left(^{\alpha}_{\beta}\right)
(0|\tau)\vartheta_1(\nu_1-\nu_2|\tau)}
\label{psipsi}
\ee
by applying the world-sheet involution $z\to 1-\bar z$
(see for instance the appendix of \cite{Antoniadis:1997vw}). Thus,
for example,
\be
\langle X(z_1)X(z_2)\rangle |_{\rm annulus}=
{1\over 2}\left(\langle X(z_1)X(z_2)\rangle
+\langle X(z_1)X(1-\bar z_2)\rangle +
\langle X(1-\bar z_1)X(z_2)\rangle+\right.
\ee
$$+\left.
\langle X(1-\bar z_1)X(1-\bar z_2)\rangle\right)\, .
$$
In eq.(\ref{psipsi}), $\alpha,\beta$ denote the fermionic spin structures
and $\vartheta$ the Jacobi theta-functions.
Setting $z_1=e^{-2\pi\nu}$ and $z_2=e^{i\pi}$ we obtain\footnote{Fixing
the second position at a different point does not affect the result.
This can be checked explicitly by shifting the integration measure.}
\begin{eqnarray}
\langle X(z_1)X(z_2)\rangle |_{\rm annulus} &=& -{1\over 2}
\log\left|{\vartheta_1(i\nu-1/2|\tau)\over
\vartheta'_1(0|\tau)}\right|^2+{\pi\nu^2\over \tau_2}\nonumber\\
&=& -{1\over 2}\log\left|{\vartheta_2(i\nu|\tau)\over
\vartheta'_1(0|\tau)}\right|^2+{\pi\nu^2\over\tau_2}\, .
\end{eqnarray}

The fermionic propagator on the torus satisfies the identity
\cite{Kiritsis:1998hj}: \be
\langle\psi(z_1)\psi(z_2)\rangle^2\left(^{\alpha}_{\beta}\right)=-{1\over
4}{\cal P}(z_1-z_2)-\pi
i\partial_{\tau}\log{\vartheta\left(^{\alpha}_{\beta}\right)(0|\tau)\over
\eta(\tau)}\, , \ee where ${\cal P}(z_1-z_2)$ is the Weierstrass
function and $\eta$ the Dedekind eta-function. The ${\cal P}$ term
as well as the scalar correlator term in eq.(\ref{ampldef}) are
spin-structure independent and their contribution vanishes upon
spin structure summation, because of space-time supersymmetry. The
rest is position independent. Dropping the ${\cal P}$-piece, we
effectively have \be
\langle\psi(z_1)\psi(z_2)\rangle^2\left(^{\alpha}_{\beta}\right)|_{\rm
annulus}= -2\pi
i\partial_{\tau}\log{\vartheta\left(^{\alpha}_{\beta}\right)(0|\tau)\over
\eta(\tau)}\, . \ee

We should mention however that we expect this result to remain valid
beyond supersymmetric vacua. In fact, in closed string threshold
calculations for gauge couplings, there is a similar expression
(see for example \cite{Kiritsis:1998hj}) and the integral over the extra
bosonic term $\langle X\partial X\rangle ^2$ cancels against the integral
over the Weierstrass function. We expect that a similar cancellation
happens also here. Extra support for this conjecture is the structure of
the open string partition function in the presence of magnetic fields,
which is also used for the calculation of threshold corrections
\cite{Antoniadis:1999ge}.

At this point we must be more explicit about the moduli integration
measure. This is given by
\be
\int [d\tau][dz]=\int_{0}^{i\infty}d\tau\int_{0}^{\tau}d(i\nu)
=-\int_{0}^{\infty}{dt\over 2}\int_{0}^{t/2}d\nu\, .
\ee
The only dependence on $\nu$ comes from the bosonic propagator:
\be
e^{-\delta
\langle X(z)X(z_0)\rangle}=
{(2\pi\eta^3(\tau))^{\delta}\over\vartheta_2(i\nu|\tau)^{\delta}
}e^{\pi\nu^2\delta\over \tau_2} =\tau_2^{\delta/2}
{(2\pi\eta^3(\tau))^{\delta}\over\vartheta_4(i\nu/\tau|-1/\tau)^{\delta}}
\, ,
\ee
where in the last step we performed a modular transformation on the
$\vartheta$-function.
We are now in position to evaluate the $\nu$ integral. Since eventually we
will set $\delta=0$ we are interested in the leading term. We obtain
\be
\int_{0}^{\tau_2}d\nu~\tau_2^{\delta/2}
{(2\pi\eta^3(\tau))^{\delta}\over \vartheta_4(i\nu/\tau|-1/\tau)^{\delta}}
=\tau_2^{1+\delta/2}
[2\pi\eta^3(\tau)]^{\delta}+\co(\delta)\, .
\label{nu}
\ee

We can now proceed to parameterize the annulus contribution to the
orientifold partition function as
\be
Z^{ab}_k={1\over 4\pi^4\tau_2^2}\sum_{\alpha,\beta=0,1}{1\over 2}
(-1)^{\alpha+\beta+\alpha\beta} {\vartheta\left[^{\alpha}_{\beta}\right]
(0|\tau)\over \eta^3(\tau)}Z^{ab}_{int,k}\left[^{\alpha}_{\beta}\right]
\, ,
\ee
where $ab$ labels the type of branes at the two endpoints of the annulus,
and $Z^{ab}_{int,k}$ is the internal part of the annulus partition
function, containing the contribution of the six compact (super)coordinates.
For the $\vartheta$-functions we use the notation and conventions of
appendix A in \cite{Kiritsis:1998hj}. In particular there are some sign
changes from the conventions of \cite{Aldazabal:1998mr}.
Putting everything together and assuming a supersymmetric ground state, we
obtain
\ba
{\cal A}^{ab}_k&=&{(2\pi)^{\delta}\over 4\pi^4|G|}
\int_0^{\infty}d\tau_2~\tau_2^{-1+\delta/2}
\eta^{3\delta}(\tau)\sum_{{\alpha,\beta=0,1},{even}}
{1\over 2}(-1)^{\alpha+\beta+\alpha\beta}
{i\pi\partial_{\tau}\vartheta\left[^{\alpha}_{\beta}\right]
(0|\tau)\over \eta^3(\tau)}Z^{ab}_{int,k}\left[^{\alpha}_{\beta}\right]
\nonumber\\
&=&{(\sqrt{2}\pi)^{\delta}\over |G|}\int_0^{\infty}dt~t^{-1+\delta/2}
\eta^{3\delta}(it/2) F^{ab}_k(t)\, ,
\label{fk}
\ea
where we have defined
\be
F_k^{ab}=\tau_2^2Z^{ab}_k=
{1\over 4\pi^4}\sum_{{\alpha,\beta=0,1},{even}}{1\over
2}(-1)^{\alpha+\beta+\alpha\beta}
{i\pi\partial_{\tau}\vartheta\left[^{\alpha}_{\beta}\right] (0|\tau)\over
\eta^3(\tau)}Z^{ab}_{int,k}\left[^{\alpha}_{\beta}\right]\, .
\ee
Note the similarity of this expression with the one appearing in the
expression of the one-loop correction to gauge couplings (see for
instance \cite{Kiritsis:1998hj} and references therein). It follows that
$F_k^{ab}$ can be formally written as a supertrace over states from the
open $ab$ $k$-orbifold sector:
\be
F_k^{ab}={|G|\over (2\pi)^2}Str^{ab}_{k, {\rm open}}\left[{1\over
12}-s^2\right]e^{-tM^2/2}\, ,
\label{fkstr}
\ee
where $s$ is the four-dimensional helicity. As we mentioned above, we
expect that this expression holds in the non supersymmetric case, as well.

\subsection{$N=1$ sectors}

In this case there is no radius dependence of the integrand. The behavior
of $F_k^{ab}$ for large $t$ is:
\be
\lim_{t\to\infty}F^{ab}_k(t)=C^{ab,IR}_k+\co(e^{-\pi t})
\ee
with
\be
C^{ab,IR}_k={|G|\over (2\pi)^2}
Str_k\left[{1\over 12}-s^2\right]_{\rm open}\, ,
\label{fk1}
\ee
where the supertrace is restricted over massless states in the open
channel $k$-sector of the orbifold. This expression is essentially the
same with the one that appears in the evaluation of the one-loop
beta-function and can be expressed in terms of the massless content of
the $k$-th sector as
\be
Str_k\left[{1\over 12}-s^2\right]=-{3\over 2}N_V+{1\over 2}N_C\, ,
\ee
where $N_{V,C}$ is the number of vector, respectively chiral multiplets
appearing in the $k$-th sector.

For small $t$ we have instead
\be
\lim_{t\to 0}F^{ab}_k(t)={1\over t}[C^{ab,UV}_k+\co(e^{-\pi/t})]
\ee
where
\be
C^{ab,UV}_k={|G|\over (2\pi)^2}Str_{k}
\left[{1\over 12}-s^2\right]_{\rm closed}
\ee
The relevant helicity supertrace
is now in the transverse closed $k$-sector mapped from the open
$k$-sector by a modular transformation. Here also, this can be
written as $-{3\over 2}N_V+{1\over 2}N_C$ where now the states are from the
closed $k$-th string sector (transverse channel).
Note that both in the direct and transverse channel all states
contribute. We should stress that this result is valid for $N=1$
sectors only. Moreover, the trivial sector $k=0$, does not contribute to
the supertrace due to its enhanced $N=4$ supersymmetry ($N_V=3N_C$).

As shown explicitly in Appendix A, the $t$-integral has a
logarithmic divergence in $\delta$ in the IR and a pole in the UV
(reflecting the UV tadpole of the anomalous $U(1)$):
\be
{\cal A}^{ab}_k= {2C^{ab,UV}_k \over \pi\delta |G|}+
{\cal O}(\log\delta)\, .
\ee
The on-shell limit can be obtained by setting $\epsilon_1=\epsilon_2$, so
that:
\be
[(\epsilon_1\cdot \epsilon_2)(p_1\cdot p_2)-(\epsilon_1\cdot
p_2)(\epsilon_2\cdot p_1)]
/(p_1\cdot p_2)\to \epsilon\cdot\epsilon
\ee
It follows that the contribution to the (unormalized) mass matrix from
$N=1$ sectors reads:
\ba
{1\over 2}M^2_{ab}|_{N=1}&=&{2\over \pi |G|}\sum_{N=1~~\rm
sectors}Tr[\gamma_k\lambda^a]Tr[\gamma_k\lambda^b] C_{k}^{ab,UV}\\
&=&{1\over 2\pi^3}\sum_{N=1~~\rm  sectors}Tr[\gamma_k\lambda^a]
Tr[\gamma_k\lambda^b]Str_{k}\left[{1\over 12}-s^2\right]_{\rm closed~~
channel}\, .\nonumber
\ea
We should remind to the reader that the multiplicities in the open
channel of the annulus partition function have a direct particle
interpretation (the projections happen at the CP factors of the
boundaries). Such an interpretation does not seem possible in the
closed string channel. Thus, our result does not seem  expressible in
terms of field theory data.

We now describe the  explicit form of this  contribution
for $Z_N$ orientifolds. The internal partition function of the
$k$-th sector is \cite{Aldazabal:1998mr}:
\be
Z^{99}_{int,k}=Z^{55}_{int,k}=\prod_{j=1}^3{(2\sin[\pi
kv_j])\vartheta\left[ {\alpha\atop
\beta+2kv_j}\right]\over\vartheta\left[ {1\atop 1-2kv_j}\right]}\, ,
\ee
\be
Z^{95}_{int,k}=-2(2\sin[\pi k
v_1]){\vartheta\left[{\alpha\atop \beta+2kv_1}\right]\over
\vartheta\left[ {1\atop
1-2kv_1}\right]}\prod_{j=2}^3{\vartheta\left[ {\alpha+1\atop
\beta+2kv_j}\right]\over\vartheta\left[ {0\atop 1-2kv_j}\right]}\, ,
\ee
where $k$ runs over the orientifold $N=1$ sectors, $(v_1,v_2,v_3)$ is
the generating rotation vector of the orbifold satisfying
$v_1+v_2+v_3=0$ in order to preserve at least $N=1$ supersymmetry and
the 5-branes are stretched along the first torus by convention. To
compare with other works, one should use the identities:
\be
\vartheta\left[ {1\atop 1+2kv_j}\right]=-\vartheta\left[
{1\atop 1-2kv_j}\right]\sp \vartheta\left[ {0\atop
1+2kv_j}\right]=\vartheta\left[ {0\atop 1-2kv_j}\right]\, .
\ee

As shown in appendix B, we can directly compute
\be
C^{99,UV}_k=C^{55,UV}_k=-{1\over 2\pi^2}\prod_{i=1}^{3}|\sin[\pi
kv_j]|\, ,
\ee
\be
C^{95,UV}_k={\sin(\pi kv_1)\over 2\pi^2}\eta_k\, ,
\ee
where
\be
\eta_k \equiv \sum_{i=1}^3\left[\{kv_i\}-{1\over 2}\right]=
{1\over 2}\prod_{i=1}^{3}{\sin[\pi kv_j] \over |\sin[\pi kv_j]|}\, .
\label{etak}
\ee
Thus, the contribution to the mass from $N=1$ sectors of $Z_N$
orbifolds is
\be
\left.\left. {1\over 2}M^2_{99,ab}\right|_{N=1}={1\over 2}M^2_{55,ab}
\right|_{N=1}=\sum_{k\atop {N=1~\rm sectors}} -{1\over \pi^3 |G|}
\prod_{i=1}^{3}|\sin[\pi kv_j]|~
Tr[\gamma_k\lambda^a]Tr[\gamma_k\lambda^b]
\label{n199}
\ee
\be
\left.{1\over 2}M^2_{95,ab}\right|_{N=1}=\sum_{k\atop {N=1~\rm sectors}}
{\sin(\pi kv_1)\over 2\pi^3|G|} \eta_k~
Tr[\gamma_k\lambda^a]Tr[\gamma_k\lambda^b]\, ,
\label{n195}
\ee
where we have divided the 59 contribution by two, to avoid overcounting.

\subsection{$N=2$ sectors}

$N=2$ sectors are present when a two-torus remains invariant under
the action of the appropriate orientifold element. Only massless
states and their KK descendants survive the helicity supertrace
(\ref{fkstr}). In this case, the function $F^{ab}_k(t)$ is given
by: \be F^{ab}_k(t)=C^{ab,IR}_k~~\Gamma_2(t)\, , \ee where
$C^{ab,IR}_k$ is still given by (\ref{fk1}). $\Gamma_2(t)$ is
either the appropriate momentum lattice when these directions are
NN (Neumann boundary conditions), or the winding lattice when
these directions are DD (Dirichlet boundary conditions)
\cite{Aldazabal:1998mr}. No lattice sum can appear along ND
directions.

For normalization purposes, the general closed string lattice sum
containing both windings and momenta can be written as
\be
Z_2=\sum_{m_i,n_i\in Z}e^{-{\pi\tau_2\alpha'\over V_2
U_2}|m_1+Um_2+T(n_1+Un_2)/\alpha'|^2-2\pi\tau_1(m_1n_1+m_2n_2)}\, ,
\label{latsum}
\ee
where $T=B+iV_2$ and $U=(G_{12}+iV_2)/G_{11}$ are, respectively, the
K\"ahler and complex structure moduli of the torus, expressed in terms of
the two-index antisymmetric tensor $B_{IJ}=B\epsilon_{IJ}$ and the
$2\times 2$ metric $G_{IJ}$ ($V_2=\sqrt{G}$). Setting the windings to
zero, and $\alpha'=1/2$, we obtain the open string momentum sum relevant
in the NN case
\be
\Gamma_2(t)=\sum_{m,n\in Z}e^{-\pi t{|m+nU|^2\over 2U_2V_2}}
={2V_2\over t}\sum_{m,n\in Z}
e^{-{2\pi V_2\over t}{|m+nU|^2\over U_2}}\, ,
\label{lattice}
\ee
while the open string (DD)
winding sum is
\be
\tilde\Gamma_2(t)=\sum_{m,n\in Z}e^{-2\pi t
V_2{|m+nU|^2\over U_2}}= {1\over 2V_2 t}\sum_{m,n\in Z} e^{-{\pi\over
2tV_2}{|m+nU|^2\over U_2}}\, .
\label{lattice1}
\ee
The normalizations above are in agreement with
\cite{Bachas:1997bp,Bachas:1996zt}. Note that the open and closed
channel supertraces (\ref{fkstr}) are now the same, since massive string
oscillator contributions cancel and one is left over with the lattice
sum (BPS states).

Using the results of appendix A, we obtain the pole contribution
\be
I^{UV}_{k}={4V_2~C^{ab,IR}_k\over \pi\delta}+{\cal O}(\log\delta)\, .
\ee
Consequently, the contribution to the mass is
\ba
\left.{1\over 2}
M^2_{ab}\right|_{N=2}&=&{4V_2\over \pi |G|}\sum_{N=2~\rm
sectors}Tr[\gamma_k\lambda^a]Tr[\gamma_k\lambda^b]C^{ab,IR}_k\\
&=&-{V_2\over \pi^3}\sum_{N=2~\rm  sectors}Tr[\gamma_k\lambda^a]
Tr[\gamma_k\lambda^b]Str_{\tilde k}\left[
{1\over 12}-s^2\right]_{\rm open~ channel}\, .\nonumber
\ea
In the DD case, relevant for mass matrix elements coming from
$D_{p<9}$ branes, the mass is similar as above with ${V_2\over
\alpha'}\to {\alpha'\over V_2}$ ($V_2\to 1/(4V_2)$ for $\alpha'=1/2$).

We now proceed to evaluate the contributions to the mass coming from
$N=2$ sectors of abelian orientifolds.
For such sectors, one of the $kv_i$ is integer.
We will choose without loss of generality $k v_1=$ integer.
The internal partition function is then
\be
Z^{99}_{int,k}=\Gamma_2{\vartheta\left[
{\alpha\atop \beta+2kv_1}\right]\over \eta^3}
\prod_{j=2}^3{(2\sin[\pi kv_j])\vartheta\left[
{\alpha\atop \beta+2kv_j}\right]\over\vartheta\left[
{1\atop 1-2kv_j}\right]}
\ee
and we can straightforwardly compute
\be
C^{ab,IR}_{k}=C^{ab,UV}_{k}={(-1)^{kv_1}\over 2\pi^2}
\prod_{j=2}^3{\sin[\pi kv_j]}=-{1\over 2\pi^2}
\prod_{j=2}^3{|\sin[\pi kv_j]|}
\ee
and
\be
\left.{1\over 2}M^2_{ab,NN}\right|_{N=2}=\sum_{k\atop {N=2~\rm sectors}}
-{2V_2\over \pi^3|G|}\prod_{j=2}^3{|\sin[\pi kv_j]|}
Tr[\gamma_k\lambda^a]Tr[\gamma_k\lambda^b]\, ,
\ee

\be
\left.{1\over 2}M^2_{ab,DD}\right|_{N=2}=\sum_{k\atop {N=2~\rm sectors}}
-{1\over 2V_2\pi^3|G|}\prod_{j=2}^3{|\sin[\pi kv_j]|}
Tr[\gamma_k\lambda^a]Tr[\gamma_k\lambda^b]\, .
\ee

Finally, for the 59 case, the relevant $N=2$ sector is when the
longitudinal torus is untwisted. In this case, the internal partition
function is given by
\be
Z^{95}_{int,k}=2\Gamma_2{\vartheta\left[
{\alpha\atop \beta+2kv_1}\right]\over \eta^3}\prod_{j=2}^3{\vartheta\left[
{\alpha+1\atop \beta+2kv_j}\right]\over\vartheta
\left[{0\atop 1-2kv_j}\right]}
\ee
and we obtain
\be
C^{95,ab,IR}_{k}=(-1)^{kv_1}{1\over 4\pi^2}
\ee
and
\be
\left.{1\over 2}M^2_{ab,DN}\right|_{N=2}=\sum_{k\atop {N=2~\rm
sectors}} (-1)^{kv_1}{V_2\over
2\pi^3|G|}Tr[\gamma_k\lambda^a]Tr[\gamma_k\lambda^b]\, .
\ee
As earlier, we have divided the 59 contribution by an additional factor
of two. In the case where the two-torus corresponds to DD boundary
conditions (in a D7-D3 configuration for instance), one should replace
$V_2\to 1/4V_2$.

\section{Explicit orientifold examples}

$N=1$, $Z_N$ orientifolds are generated by a rotation that acts as
\be g~X^i=e^{2\pi iv_i} X^i\;\;\;,\;\;\;g~\bar X^i= e^{-2\pi i
v_i}\bar X^i\, , \ee where $X^i,\bar X^i$ are the complex
coordinates of the the three two-tori. The parameters $v_i$
determining the fundamental $Z_N$ rotation satisfy $Nv_i\in Z$ and
$v_1+v_2+v_3=0$ in order to preserve space-time supersymmetry.

The action on the Chan--Patton indices is determined by the matrices
$\gamma_k=(\gamma_1)^k$ representing the action of the orbifold element
$g^k$, as
\be
\gamma_{k}=e^{-2\pi i{\hat v}\cdot H}\, ,
\ee
where $H_I, I=1,\cdots,16$ are the Cartan generators of $SO(32)$ and
${\hat v}^I$ is a rational vector specific to any given orbifold. A basis
for the
Cartan generators is given by diagonal matrices having the $\sigma^3$
Pauli matrix somewhere in the diagonal and zero everywhere else (so that
$Tr[H_I^2]=2$). There is a vector ${\hat v}_9$ for D9-branes and different
vectors (${\hat v}_5$) for every potential set of D5-branes.

\subsection{The $Z_3$ orientifold}

Here there are no D5-branes \cite{Angelantonj:1996uy}. The
orbifold rotation vector is $(v_1,v_2,v_3)=(1,1,-2)/3$ and the
Chan--Patton projection vector is \be {\hat v}_9={1\over
3}(1,1,1,1,1,1,1,1,1,1,1,1,0,0,0,0) \ee with
$Tr[\gamma_1]=Tr[\gamma_2]=-4$. This breaks $SO(32)$ to
$U(12)\times SO(8)$. The $U(1)$ factor of $U(12)$ is anomalous.
The normalized generator of the anomalous $U(1)$ is \be
\lambda={1\over 4\sqrt{3}}\sum_{i=1}^{12}H_I\sp
tr[\lambda^2]={1\over 2}\, . \ee Thus, we can compute \be
Tr[\gamma_1\lambda]=-2\sqrt{3}i\sin{(2\pi/3)}=-3i\sp
Tr[\gamma_2\lambda]=-2\sqrt{3}i\sin{(4\pi/3)}=3i\, , \ee \be
Tr[\gamma_1\lambda^2]=\cos{(2\pi/3)}=-{1\over 2} \sp
Tr[\gamma_2\lambda^2]=\cos{(4\pi/3)}=-{1\over 2}\, . \ee Using
(\ref{n199}), we can now evaluate the anomalous gauge boson mass:
\be {1\over 2}M^2=-{1\over 3\pi^3}\left[\sin^3(\pi/3)
Tr[\gamma_1\lambda]^2 +\sin^3(2\pi/3)
Tr[\gamma_2\lambda]^2\right]={9\sqrt{3}\over 4\pi^3}\, . \ee
Putting back $M^2_s=1/\alpha'$ from the $2\alpha'=1$ convention
and taking into account the normalization of the $F^2$ kinetic
terms $2Tr[\lambda^2]/4g_A^2$, we obtain for the normalized gauge
boson mass \be M_{phys}^2={9\sqrt{3}\over 4\pi^3}g_A^2~M_s^2\, .
\label{M99} \ee

Note that this example can be used to realize two out of the four
possible configurations for the abelian gauge bosons and their
corresponding axions, displayed in eq.(\ref{cases}), namely the cases
(brane, brane) and (bulk, brane). Indeed, the RR axions from the twisted
closed string sector are localized in all six internal dimensions, while
the anomalous $U(1)$ can be either in the bulk (on the D9-branes), or on
the brane with respect to directions that are T-dualized, so that one has
D$p$-branes with $p<9$. Moreover, the $U(1)$ gauge coupling in
eq.(\ref{M99}) is given in general by $g_A^2=g_s{V}_\parallel$, with
${V}_\parallel$ the internal volume (in string units) of the $p-3$
compactified directions along the D$p$-brane.

\subsection{The $Z_7$ orientifold}

The orbifold rotation vector is $(v_1,v_2,v_3)=(1,2,-3)/7$.
Tadpole cancellation implies the existence of 32 D9-branes. The
Chan--Patton vector is \be {\hat v}_9={1\over
7}(1,1,1,1,2,2,2,2,-3,-3,-3,-3,0,0,0,0) \ee which implies \be
Tr[\gamma_k]=4\sp k=1,2,3,4,5,6\, . \ee The gauge group is
$U(4)^3\times SO(8)$ and there are only $N=1$ sectors.

The potentially anomalous $U(1)$s are the abelian factors of the gauge
group and the relevant CP matrices are
\be
\lambda_1={1\over 4}\sum_{I=1}^{4} H_I\sp
\lambda_2={1\over 4} \sum_{I=5}^{8} H_I\sp
\lambda_3={1\over 4}\sum_{I=9}^{12}H_I\, ,
\ee
which satisfy $tr[\lambda_i\lambda_j]={1\over 2}\delta_{ij}$.
The four-dimensional mixed non-abelian anomalies of these $U(1)$s
are proportional to the matrix
\be
\left(\matrix{2&0&-4&\cr -4&2&0\cr 0&-4&2\cr 4&4&4\cr }\right)\, ,
\ee
where the columns label the $U(1)$s while the rows label the
non-abelian factors $SU(4)^3\times SO(8)$. It follows that all three
$U(1)$s are anomalous.
We also have $\eta_1=\eta_2=-\eta_3=\eta_4=-\eta_5=-\eta_6=-1/2$ (see
eq.(\ref{etak})). The contributions to the mass matrix are:
\be
{1\over 2}M^2_{ij}=-{\sin{\pi\over 7}\sin{2\pi\over 7}
\sin{3\pi\over 7}  \over
7\pi^3}\sum_{k=1}^{6}Tr[\gamma_k\lambda_i]Tr[\gamma_k\lambda_j]=
2{\sin{\pi\over 7}\sin{2\pi\over 7} \sin{3\pi\over 7}  \over
7\pi^3}\delta_{ij}
\ee
and there is no mixing in this case.

\subsection{The $Z'_6$ orientifold}

The orbifold rotation vector is $(v_1,v_2,v_3)=(1,-3,2)/6$. There
is an order two twist ($k=3$) and we must have one set of
D5-branes. Tadpole cancellation then implies the existence of 32
D9-branes and 32 D5-branes that we put together at one of the
fixed points of the $Z_2$ action (say the origin). The
Chan--Patton vectors are \be {\hat v}_9={\hat v}_5={1\over
12}(1,1,1,1,5,5,5,5,3,3,3,3,3,3,3,3) \ee which imply \be
Tr[\gamma_k]=0\sp k=1,3,5\sp Tr[\gamma_2]=-8\sp Tr[\gamma_4]=8\, .
\ee The gauge group has a factor of $U(4)\times U(4)\times U(8)$
coming from the D9-branes and an isomorphic factor coming from the
D5-branes. The $N=1$ sectors correspond to $k=1,5$, while for
$k=2,3,4$ we have $N=2$ sectors.

The potentially anomalous $U(1)$s are the abelian factors of the gauge
group and the relevant CP matrices for the D9-branes are:
\be
\lambda_1={1\over 4}\sum_{I=1}^{4} H_I\sp
\lambda_2={1\over 4}\sum_{I=5}^{8} H_I\sp
\lambda_3={1\over 4\sqrt{2}}\sum_{I=9}^{16} H_I\, ,
\ee
which satisfy $tr[\lambda_i\lambda_j]={1\over 2}\delta_{ij}$.
Similar formulae apply to the other three $U(1)$ matrices
$\tilde\lambda_i$ coming from the D5-sector. The four-dimensional
anomalies of these $U(1)$s (and their cancellation mechanism) were
computed in \cite{Ibanez:1999qp}. The mixed anomalies with the six
non-abelian groups are given by the matrix\footnote{Note that here
we use a different normalization for the $U(1)$ generators than in
\cite{Ibanez:1999qp}.}
\be
\left(\matrix{ 2&2&4\sqrt{2}&-2&0&-2\sqrt{2}\cr
-2&-2&-4\sqrt{2}&0&2&2\sqrt{2}\cr 0&0&0&2&-2&0\cr
-2&0&-2\sqrt{2}&2&2&4\sqrt{2}\cr 0&2&2\sqrt{2}&-2&-2&-4\sqrt{2}\cr
2&-2&0&0&0&0\cr}\right)\, ,
\ee
where the columns label the $U(1)$s while the
rows label the non-abelian factors $SU(4)_9^2\times SU(8)_9\times
SU(4)^2_5\times SU(8)_5$. The upper 3$\times$3 part corresponds
to the 99 sector and the lower one to the 55 sector. As can be seen
by this matrix, the two linear combinations $\sqrt{2}(A_1+A_2)-A_3$
and $\sqrt{2}(\tilde A_1+\tilde A_2)-\tilde A_3$ are free of mixed
non-abelian anomalies. It can also be shown that they are also
free of mixed $U(1)$ anomalies.
We can now compute:
\be
Tr[\gamma_k\lambda_1]=-2i\sin\left({\pi k\over 6}\right),\
Tr[\gamma_k\lambda_2]=-2i\sin\left({5\pi k\over 6}\right),\
Tr[\gamma_k\lambda_3]=-2i\sqrt{2}\sin\left({\pi k\over 2}\right)
\ee
\be
Tr[\gamma_k\tilde\lambda_1]=-2i\sin\left({\pi k\over 6}\right),\
Tr[\gamma_k\tilde\lambda_2]=-2i\sin\left({5\pi k\over 6}\right),\
Tr[\gamma_k\tilde\lambda_3]=-2i\sqrt{2}\sin\left({\pi k\over 2}\right)
\ee
\be
Tr[\gamma_k\lambda_1^2]={1\over 2}\cos\left({\pi k\over 6}\right)\spa
Tr[\gamma_k\lambda_2^2]={1\over 2}\cos\left({5\pi k\over 6}\right)\spa
Tr[\gamma_k\lambda_3^2]={1\over 2}\cos\left({\pi k\over 2}\right)
\ee
while $Tr[\gamma_k\lambda_i\lambda_j]=0$ for $i\not= j$. We
also have $\eta_1=\eta_2=\eta_4=-\eta_5=-1/2$.

The contribution to the mass matrix from $N=1$ sectors is:
\be
{1\over 2}M^2_{99,ij}=-{\sqrt{3}\over 24\pi^3}\left(Tr[\gamma_1\lambda_i]
Tr[\gamma_1\lambda_j]+Tr[\gamma_5\lambda_i]Tr[\gamma_5\lambda_j]\right)
\ee
and similarly for $M_{55,ij}$, while
\ba
{1\over 2}M^2_{95,ij}
=&-&{\sqrt{3}\over 48\pi^3}\left(Tr[\gamma_1\lambda_i]
Tr[\gamma_1\tilde\lambda_j]+Tr[\gamma_5\lambda_i]
Tr[\gamma_5\tilde\lambda_j]\right. \\
&+& \left. Tr[\gamma_2\lambda_i]Tr[\gamma_2\tilde\lambda_j]-
Tr[\gamma_4\lambda_i]Tr[\gamma_4\tilde\lambda_j]\right)\, . \nonumber
\ea
On the other hand, the contributions from $N=2$ sectors read:
\ba
{1\over 2}M^2_{99,ij}=&-&{V_2\over 4\pi^3}\left(Tr[\gamma_2\lambda_i]
Tr[\gamma_2\lambda_j]+Tr[\gamma_4\lambda_i]Tr[\gamma_4\lambda_j]\right)\\
&-&{V_3\over 3\pi^3}Tr[\gamma_3\lambda_i]Tr[\gamma_3\lambda_j]\nonumber
\ea
\ba
{1\over 2}M^2_{55,ij}=&-&{1\over
16V_2\pi^3}\left(Tr[\gamma_2\tilde\lambda_i]Tr[\gamma_2\tilde\lambda_j]+
Tr[\gamma_4\tilde\lambda_i]Tr[\gamma_4\tilde\lambda_j]\right)\\
&-&{V_3\over 3\pi^3}Tr[\gamma_3\lambda_i]Tr[\gamma_3\lambda_j]\nonumber
\ea
and
\be
{1\over 2}M^2_{95,ij}=- {V_3\over
12\pi^3}Tr[\gamma_3\lambda_i]Tr[\gamma_3\tilde\lambda_j]\, .
\ee
Thus, the unormalized mass matrix has eigenvalues and eigenvectors:
\be
m_1^2=6V_2\sp -A_1+A_2\, ;
\label{bubu}
\ee
\be
m_2^2={3\over 2V_2}\sp -\tilde A_1+\tilde A_2\, ;
\label{brbu}
\ee
\be
m_{3,4}^2={5\sqrt{3}+48V_3\pm\sqrt{3(25-128
\sqrt{3}V_3+768V_3^2)}\over 12}\, ,
\ee
with respective eigenvectors
\be
\pm a_{\pm}(A_1+A_2-\tilde A_1-\tilde A_2)-A_3+\tilde A_3
\ee
where
\be a_{\pm}={\mp 3+\sqrt{25-128\sqrt{3}V_3+768V_3^2}\over 4\sqrt{2}
(4\sqrt{3}V_3-1)}\, ;
\ee
\be
m_{5,6}^2={15\sqrt{3}+80V_3\pm\sqrt{5(135-384\sqrt{3}V_3+1280V_3^2)}
\over 12}\, ,
\ee
with respective eigenvectors
\be
\pm
b_{\pm}(A_1+A_2+\tilde A_1+\tilde A_2)+A_3+\tilde A_3
\ee
where
\be
b_{\pm}={\pm 9\sqrt{3}-\sqrt{5(135-384\sqrt{3}V_3+1280V_3^2)}
\over 4\sqrt{2}(20V_3-3\sqrt{3})}\, .
\ee

Note that the eigenvalues are always positive. They are also
invariant under the T-duality symmetry of the theory $V_2\to 1/4V_2$.
Thus, all $U(1)$s become massive, including the two anomaly free
combinations. The reason is that these combinations are anomalous in
six dimensions. Observe however that in the limit $V_3\to 0$, the two
linear combinations that are free of four-dimensional anomalies
become massless. This is consistent with the fact that the
six-dimensional anomalies responsible for their mass cancel locally
in this limit.

To obtain the normalized mass matrix, we must also take into account the
kinetic terms of the $U(1)$ gauge bosons which are \be S_{\rm
kinetic}=-{1\over
4g_s}\left[V_1V_2V_3(F_1^2+F_2^2+F_3^2) +V_3(\tilde F_1^2+\tilde
F_2^2+\tilde
F_3^2)\right]\, . \ee This implies $M^2_{99}\to M^2_{99}/(V_1V_2V_3)$,
$M^2_{55}\to M^2_{55}/V_3$ and $M^2_{95}\to M^2_{95}/(\sqrt{V_1V_2}V_3)$.
The
resulting eigenvalues are too complicated and not illuminating to produce
here.

Strictly speaking the formulae we presented should be used for $V_3\geq 1$.
When $V_3<1$ we can T-dualize and rewrite the theory in terms of D3-D7
branes.
Then the unormalized mass remains as above with $V_3\to 1/4V_3$ but the
kinetic
terms of the gauge bosons are no longer multiplied by $V_3$.

Using the $Z'_6$ orientifold, one can realize the remaining two
possible configurations for the anomalous $U(1)$ gauge fields and
their corresponding axions, namely the (bulk, bulk) and (brane, bulk)
cases of eq.(\ref{cases}); the other two were realized for instance
in the context of $Z_3$ orientifold, as we described before. In fact,
identifying the second torus with the bulk, the two configurations
correspond to the cases (\ref{bubu}) and (\ref{brbu}), respectively,
that receive contributions from the corresponding $N=2$ sector only.

\subsection{The $Z_6$ orientifold}

The orbifold rotation vector is $(v_1,v_2,v_3)=(1,1,-2)/6$. There
is an order two twist ($k=3$) and we must have one set of
D5-branes. Tadpole cancellation then implies the existence of 32
D9-branes and 32 D5-branes, as in the previous example, that we
put together at the origin of the internal space. The Chan--Patton
vectors are \be {\hat v}_9={\hat v}_5={1\over
12}(1,1,1,1,1,1,5,5,5,5,5,5,3,3,3,3) \ee implying \be
Tr[\gamma_k]=0\ \ \ {\rm for}\ \ k=1,3,5\sp Tr[\gamma_2]=4\sp
Tr[\gamma_4]=-4\, . \ee The gauge group has a factor of
$U(6)\times U(6)\times U(4)$ coming from the D9-branes and an
isomorphic factor coming from the D5-branes. The $N=1$ sectors
correspond to $k=1,2,4,5$, while $k=3$ is an $N=2$ sector.

The potentially anomalous $U(1)$s are the abelian factors of the
gauge group and the relevant CP matrices for the D9-branes are:
\be
\lambda_1={1\over 2\sqrt{6}}\sum_{I=1}^{6} H_I\sp
\lambda_2={1\over 2\sqrt{6}}
\sum_{I=7}^{12} H_I\sp \lambda_3={1\over 4}\sum_{I=13}^{16} H_I\, ,
\ee
that satisfy $tr[\lambda_i\lambda_j]={1\over 2}\delta_{ij}$.
Similar formulae apply to the other three $U(1)$ matrices
$\tilde\lambda_i$ coming from the D5-sector.
The four-dimensional mixed non-abelian anomalies of these $U(1)$s
 are proportional to the matrix
\be
\left(\matrix{6&-3&\sqrt{6}&3&0&\sqrt{6}\cr
3&-6&-\sqrt{6}&0&-3&-\sqrt{6}\cr
-9&9&0&-3&3&0\cr
3&0&\sqrt{6}&6&-3&\sqrt{6}\cr 0&-3&-\sqrt{6}&3&-6&-\sqrt{6}\cr
-3&23&0&-9&9&0\cr}\right)\, .
\ee
The columns label the $U(1)$s, while the rows label the non-abelian
factors $SU(6)_9^2\times SU(4)_9\times SU(6)^2_5\times SU(4)_5$. The
upper 3$\times$3 part corresponds to the 99 sector and the lower one
to the 55 sector. As can be seen by this matrix, there are three
linear combinations $A_1+A_2-\sqrt{3\over 2}A_3$, $\tilde A_1+\tilde
A_2-\sqrt{3\over 2}\tilde A_3$ and $A_3-\tilde A_3$ that are free of
mixed non-abelian anomalies. It can be shown that they are also
free of mixed $U(1)$ anomalies.

We can now compute
\be
Tr[\gamma_k\lambda_1]=-i\sqrt{6}\sin{\pi k\over 6}\spa
Tr[\gamma_k\lambda_2]=(-1)^ki\sqrt{6}\sin{\pi k\over 6}\spa
Tr[\gamma_k\lambda_3]=-2i\sin{\pi k\over 2}
\ee
and similarly for $\tilde \lambda i$. Also
\be
Tr[\gamma_k\lambda_1^2]={1\over 4}\cos{\pi k\over 6}\spa
Tr[\gamma_k\lambda_2^2]={(-1)^k\over 4}\cos{\pi k\over 6}\spa
Tr[\gamma_k\lambda_3^2]={1\over 4}\cos{\pi k\over 2}\, ,
\ee
while $Tr[\gamma_k\lambda_i\lambda_j]=0$ for $i\not= j$. Finally
$\eta_1=\eta_2=\eta_3=-\eta_4=-\eta_5=-1/2$.

The various contributions to the mass matrix are
\be
{1\over 2}M^2_{99,ij}=-{\sqrt{3}\over 48\pi^3}
\left[Tr[\gamma_1\lambda_i]Tr[\gamma_1\lambda_j]+Tr[\gamma_5\lambda_i]
Tr[\gamma_5\lambda_j]\right.
\ee
$$
\left.+3(Tr[\gamma_2\lambda_i]Tr[\gamma_2\lambda_j]+
Tr[\gamma_4\lambda_i]Tr[\gamma_4\lambda_j])\right]
-{V_3\over 3\pi^3}Tr[\gamma_3\lambda_i]Tr[\gamma_3\lambda_j]
$$
and similarly for $M_{55,ij}$, while
\be
{1\over 2}M^2_{95,ij}=-{\sqrt{3}\over 48\pi^3}\left(
[Tr[\gamma_1\lambda_i]Tr[\gamma_1\lambda_j]+
Tr[\gamma_5\lambda_i]Tr[\gamma_5\lambda_j]\right.
\ee
$$
\left.+Tr[\gamma_2\lambda_i]Tr[\gamma_2\lambda_j]
+Tr[\gamma_4\lambda_i]Tr[\gamma_4\lambda_j]\right)
-{V_3\over 12\pi^3}Tr[\gamma_3\lambda_i]Tr[\gamma_3\lambda_j]\, .
$$
This mass matrix has the following eigenvalues and eigenvectors:
\be
m_1^2=0\sp A_1+A_2-\tilde A_1-\tilde A_2+\sqrt{6}(A_3-\tilde A_3)\, ;
\ee
\be
m_2^2={3\sqrt{3}\over 2}\sp A_1-A_2-\tilde A_1+\tilde A_2\, ;
\ee
\be
m_3^2={3\sqrt{3}}\sp A_1-A_2+\tilde A_1-\tilde A_2\, ;
\ee
\be
m_4^2={40\over 3}V_3\sp -\sqrt{3\over 2}(A_1+A_2-\tilde A_1-\tilde
A_2)-A_3+\tilde A_3\, ;
\ee
\be
m^2_{\pm}={7\sqrt{3}+80V_3\pm\sqrt{147-1040\sqrt{3}V_3+6400V_3^2}
\over 12}\sp a_{\pm}(A_1+A_2+\tilde A_1+\tilde A_2)+A_3+\tilde A_3
\ee
with
\be
a_{\pm}={40V_3-\sqrt{3}\pm\sqrt{147-1040\sqrt{3}V_3+6400V_3^2}\over
12\sqrt{2}-40\sqrt{6}V_3}\, .
\ee
In the limit $V_3\to 0$ two more masses become zero ($m_4$ and $m_-$).
It is straightforward to check that the appropriate linear
combinations of $U(1)$s are anomaly-free in four dimensions.

\subsection{The $Z_3\times Z_6$ orientifold}

The orbifold rotation vectors are $v_{\theta}=(1,0,-1)/3$ and
$v_{h}=(1,-1,0)/6$. There is an order two twist $h^3$. Tadpole
cancellation implies the existence of 32 D9-branes and 32
D5-branes that we put together at the origin of the internal
space. The Chan--Patton vectors are \be {\hat v}^{\theta}_9={\hat
v}^{\theta}_5 ={1\over 3}(2,2,0,0,1,1,0,0,1,1,2,2,0,0,0,0) \ee and
\be {\hat v}^{h}_9={\hat v}^{h}_5= {1\over
12}(1,1,1,1,5,5,5,5,3,3,3,3,3,3,3,3)\, . \ee The gauge group has a
factor of $U(2)^6\times U(4)$ coming from the D9-branes and an
isomorphic factor coming from the D5-branes. Sectors are labelled
by the group elements $\theta^k h^l$. The $N=2$ sectors in the 99
and 55 configurations are
$(k,l)\in\{(1,0),(2,0),(0,1),(0,2),(2,2),(0,3)$,
$(0,4),(1,4),(0,5)\}$. In the 95 configuration we have fewer $N=2$
sectors, namely $(k,l)\in\{(0,1),(0,2),(0,3),(0,4),(0,5)\}$.

The potentially anomalous $U(1)$s are the fourteen abelian factors of the
gauge
group and the relevant CP matrices for the D9-branes are \be
\lambda_1={1\over
2}\sum_{I=1}^{2} H_I\sp \lambda_2={1\over 2}\sum_{I=3}^{4} H_I\sp
\lambda_3={1\over 2}\sum_{I=5}^{6} H_I\sp \lambda_4={1\over 2}\sum_{I=7}^{8}
H_I\, , \ee \be \lambda_5={1\over 2}\sum_{I=9}^{10} H_I\sp \lambda_6={1\over
2}\sum_{I=11}^{12} H_I\sp \lambda_7={1\over 2}\sum_{I=13}^{16} H_I\, . \ee
Similar formulae apply to the other seven $U(1)$ matrices $\tilde\lambda_i$
coming from the D5-sector. The four-dimensional mixed non-abelian anomalies
of
these $U(1)$s are proportional to the matrix \be \left(\matrix{ 1 & -1 & 0 &
0
& 0 & -1 & 2 & 0 & 0 & 0 & 0 & 0 & 0 & 0 \cr 1 & -1 & 0 & 0 & 1 & 0 & -2 & 0
&
0 & 0 & 0 & 0 & 0 & 0 \cr 0 & 0 & -1 & 1 & 1 & 0 & -2 & 0 & 0 & 0 & 0 & 0 &
0 &
0 \cr 0 & 0 & -1 & 1 & 0 & -1 & 2 & 0 & 0 & 0 & 0 & 0 & 0 & 0 \cr 0 & -1
& -1 &
0 & 0 & -1 & 2 & 0 & 0 & 0 & 0 & 0 & 0 & 0 \cr 1 & 0 & 0 & 1 & 1 & 0 & -2 &
0 &
0 & 0 & 0 & 0 & 0 & 0 \cr -1 & 1 & 1 & -1 & -1 & 1 & 0 & 0 & 0 & 0 & 0 & 0 &
0
& 0 \cr 0 & 0 & 0 & 0 & 0 & 0 & 0 & 1 & -1 & 0 & 0 & 0 & -1 & 2 \cr 0 & 0 &
0 &
0 & 0 & 0 & 0 & 1 & -1 & 0 & 0 & 1 & 0 & -2 \cr 0 & 0 & 0 & 0 & 0 & 0 & 0 &
0 &
0 & -1 & 1 & 1 & 0 & -2 \cr 0 & 0 & 0 & 0 & 0 & 0 & 0 & 0 & 0 & -1 & 1 & 0
& -1
& 2 \cr 0 & 0 & 0 & 0 & 0 & 0 & 0 & 0 & -1 & -1 & 0 & 0 & -1 & 2 \cr 0 & 0 &
0
& 0 & 0 & 0 & 0 & 1 & 0 & 0 & 1 & 1 & 0 & -2 \cr 0 & 0 & 0 & 0 & 0 & 0 & 0
& -1
& 1 & 1 & -1 & -1 & 1 & 0 \cr }\right) \ee The columns label the $U(1)$s
while
the rows label the non-abelian factors $SU(2)_9^6\times SU(4)_9\times
SU(2)^6_5\times SU(4)_5$. The upper 7$\times$7 part corresponds to the 99
sector and the lower one to the 55 sector. As can be seen by this matrix,
there
are six linear combinations \be A_1-A_3-A_5+A_6\sp A_2-A_4+A_5-A_6\sp
2(A_5+A_6)+A_7 \label{s1}\ee \be \tilde A_1-\tilde A_3+\tilde A_5-\tilde
A_6\sp
\tilde A_2-\tilde A_4+\tilde A_5-\tilde A_6\sp 2(\tilde A_5+\tilde
A_6)+\tilde
A_7 \label{s2}\ee that are free of mixed non-abelian anomalies. Mixed $U(1)$
anomalies also cancel. We can also compute: \be
\eta_{(1,1)}=\eta_{(2,1)}=\eta_{(1,2)}=\eta_{(1,3)}=-\eta_{(2,3)}=
-\eta_{(2,4)}=-\eta_{(1,5)}=-\eta_{(2,5)}={1\over 2}\, , \ee
$$
\eta_{(2,2)}=\eta_{(1,4)}=\eta_{(1,0)}=\eta_{(2,0)}=0\, .
$$

The mass matrix is given by
\ba
{1\over 2}M^2_{99,ij}=&-&\sum_{k,l\atop N=1 ~{\rm sectors}}
{s[k,l]\over 18\pi^3}Tr[\gamma_{k,l}\lambda_i]
Tr[\gamma_{k,l}\lambda_j]-{V_3\over 9}\sum_{l=1}^{5}\sin^2
\left[{\pi l\over 6}\right]Tr[\gamma_{0,l}\lambda_i]
Tr[\gamma_{0,l}\lambda_j]\nonumber\\
&-&{V_1\over 9}\sin\left[{\pi\over 3}\right]\sin
\left[{2\pi\over 3}\right](Tr[\gamma_{2,2}\lambda_i]
Tr[\gamma_{2,2}\lambda_j]+Tr[\gamma_{1,4}\lambda_i]
Tr[\gamma_{1,4}\lambda_j])\nonumber\\
&-&{V_2\over 9}\sum_{k=1}^{2}\sin^2\left[{\pi k\over
3}\right]Tr[\gamma_{k,0}\lambda_i]Tr[\gamma_{k,0}\lambda_j]\, ,
\ea
where
\be
s[k,l]\equiv \left|\sin\left[\pi\left({2k+l\over 6}\right)\right]
\sin\left[\pi{l \over 6}\right]\sin\left[\pi{k\over 3}\right]\right|
\, ,
\ee
and similarly for $M_{55,ij}$ with $V_1\to 1/4V_1$, $V_2\to 1/4V_2$,
while
\be
{1\over 2}M^2_{95,ij}=\sum_{k,l\atop N=1~ {\rm sectors}}
{\eta_{k,l} \over 36\pi^3}\sin\left[{\pi k\over 3}\right]
Tr[\gamma_{k,l}\lambda_i]Tr[\gamma_{k,l}\lambda_j]
\ee
$$
+{V_3\over 36}\sum_{l=1}^{5}\sin^2\left[{\pi l\over 6}\right]
Tr[\gamma_{0,l}\lambda_i]Tr[\gamma_{0,l}\lambda_j]\, .
$$
It follows that there are no massless gauge bosons. The mass-squared matrix
has
a double eigenvalue $4\sqrt{3}$ and a double eigenvalue $6\sqrt{3}$. It has
six
eigenvalues that depend on $V_3$ and the rest depend on all three internal
volumes. At $V_3=0$ there are two zero  eigenvalues corresponding to the
last
linear combinations in (\ref{s1},\ref{s2}), a double eigenvalue $4\sqrt{3}$
and
a double eigenvalue $6\sqrt{3}$, double eigenvalues
$(49\sqrt{3}\pm\sqrt{5259})/18$ and the rest are
\be
4(V_1+V_2\pm\sqrt{V_1^2+V_2^2-V_1V_2})
\ee
with eigenvectors purely on the
D9-branes and their duals with eigenvectors only on the D5-branes.

\section{Conclusions}

In this work we did an explicit one-loop string computation of the
$U(1)$ masses in four-dimensional orientifolds and studied their
localization properties in the internal compactified space. We have
shown that non vanishing mass-terms appear for all $U(1)$s
that are anomalous in four dimensions, but also for apparent
anomaly free combinations if they acquire anomalies in a
six-dimensional decompactification limit. In both cases, the global
$U(1)$ symmetry remains unbroken at the orientifold point, to all
orders in perturbation theory.

For supersymmetric compactifications, we found that $N=1$ sectors
lead to contributions to $U(1)$ masses that are localized in all six
internal dimensions, while those of $N=2$ sectors are localized only
in four internal dimensions. All these mass terms are described
as Green-Schwarz couplings involving axions coming from the RR
closed string sector, that transform under the corresponding $U(1)$
gauge transformations. One can thus provide explicit realizations in
brane world models of all possible configurations (\ref{cases}) for
the gauge field and the axion, propagating in the bulk of large extra
dimensions, or being localized on a brane. $N=1$ sectors describe
axions localized on a 3-brane, while $N=2$ sectors describe axions
propagating in two extra dimensions.

Our results can in principle easily be generalized to non
supersymmetric orientifolds. A particularly interesting class of
non supersymmetric constructions is given in the context of
``brane supersymmetry breaking", where supersymmetry is broken
only in the open string sector while it remains exact (to lowest
order) in the closed string bulk \cite{Antoniadis:1999xk}. In the
simplest case, the breaking of supersymmetry arises only from
combinations of D-branes with (anti)-orientifold planes which
affect only the M\"obius amplitude and thus do not change the
expression for the mass. Indeed, the latter appears as a contact
term of the annulus that remains supersymmetric. On the other
hand, in the case where the supersymmetry breaking arises also
from configurations of branes with anti-branes, there is an
additional contribution to the mass that can be easily computed
following our general method.

Our analysis has direct implications for model building
\cite{akrt}. In particular, special care is needed to guarantee
that the $U(1)$ hypercharge remains massless despite the fact that
it is anomaly free. An additional condition should be satisfied,
namely that it remains anomaly free in any six-dimensional
decompactification limit. On the other hand, anomalous $U(1)$s
could be used to reduce the rank of the low-energy gauge group and
guarantee the conservation of global symmetries, such as the
baryon and lepton number. Finally, the associated $U(1)$ gauge
bosons could be produced in particle accelerators with new
interesting experimental signals. Their masses are always lighter
than the string scale, varying from a loop factor to a much bigger
suppression by the volume of the bulk, giving rise to possible new
(repulsive) forces at sub-millimeter distances, much stronger than
gravity.


\section*{Acknowledgments}

We thank C. Angelantonj and T. Tomaras for discussions. I.A. would
like to acknowledge early collaboration with S. Dimopoulos and J.
March--Russel on gauge bosons and axions in the bulk and baryon
number conservation. This work was partially supported by the
European Commission under the RTN contracts HPRN--CT--2000--00148,
HPRN--CT--2000--00122, HPRN--CT--2000--00131, HPRN-CT-2000-00152,
the INTAS contract N 99 0590, and by the Marie Curie contract
MCFI-2001-0214.


\appendixA{Appendix A: Ultraviolet poles and Infrared logarithms}

In this appendix we calculate the UV tadpole (pole in $\delta$).
To this end, we split the integral of eq.(\ref{fk}) into UV and IR
parts:
\be
{\cal A}_k^{ab}=I^{ab,IR}_{k}+I^{ab,UV}_{k}\, .
\ee
We will first consider $N=1$ sectors, where no lattice sum appears in
the internal partition function. The behavior in the IR is:
\ba
I^{ab,IR}_{k}&=&{(\sqrt{2}\pi)^{\delta}\over |G|}
\int_1^{\infty}dt~t^{-1+\delta/2} \eta^{3\delta}(it/2)F^{ab}_k(t)\\
&=&{(\sqrt{2}\pi)^{\delta}C^{ab,IR}_k\over |G|}
\int_1^{\infty}dt~t^{-1+\delta/2}e^{-{\pi t\delta\over 8}}
+{\rm finite}\nonumber
\ea
Changing variables, we obtain
\be
I^{ab,IR}_{k}=\left({16\pi\over \delta}\right)^{\delta\over 2}
{C^{ab,IR}_k\over |G|}\int_{\pi\delta/8}^{\infty}
du~u^{-1+\delta/2}e^{-u}+{\rm finite}=\left({16\pi\over \delta}
\right)^{\delta\over 2}
{C^{ab,IR}_k\over |G|}\Gamma(\delta/2,\pi\delta/8)\, ,
\ee
where
$\Gamma(a,x)$ is the incomplete $\Gamma$-function with asymptotic
expansion for small argument x:
\be
\Gamma(a,x)=\Gamma(a)-{x^a\over a}-
\sum_{n=1}^{\infty}{(-1)^nx^{a+n}\over n!(n+a)}\, .
\ee
We thus obtain
\be
I^{ab,IR}_{k}=-{C^{ab,IR}_k\over |G|}\log{\pi\delta\over 8}
+{\rm finite}
\ee
To study the UV behavior, we use $\eta(it/2)=(t/2)^{-1/2}\eta(2/t)$
and consider
\ba
I^{ab,UV}_{k}&=&{(\sqrt{2}\pi)^{\delta}\over |G|}
\int_0^{1}dt~t^{-1+\delta/2} \eta^{3\delta}(it/2)
F^{ab}_k(t)\nonumber\\
&=&{(4\pi)^{\delta}C^{ab,UV}_k\over |G|}\int_0^{1}
dt~t^{-2-\delta}~e^{-\pi\delta/2t}+{\rm finite}\\
&=&{C^{ab,UV}_k\over |G|} \left({8\over \delta}\right)^{\delta}
{2\over \pi\delta}
           \Gamma(\delta+1,\pi\delta/2)+{\rm
finite}={2C^{ab,UV}_k \over \pi\delta |G|}+{\rm finite}\, ,\nonumber
\ea
leading to the pole, as advertised.

We will now focus on the $N=2$ sectors. Here
\be
F^{ab}_k(t)=C^{ab,IR}_k~~\Gamma_2(t)\, ,
\ee
where $C^{ab,IR}_k$ is given by (\ref{fk1}). The lattice sum is
given by (\ref{lattice}) in the NN case, and by (\ref{lattice1}) in the DD
case. To obtain the UV contribution, we have to use the second form of
the lattice sums in (\ref{lattice}) and (\ref{lattice1}). We then
find:
\ba
I^{UV}_{k}&=&(4\pi)^{\delta}C^{ab,IR}_k\int_0^{1}
dt~t^{-2-\delta}~e^{-\pi\delta/2t}\Gamma_2(t)+{\rm finite}\nonumber\\
&=&{2V_2}~C^{ab,IR}_k (4\pi)^{\delta}\left({2\over\pi\delta}
\right)^{\delta+1}\Gamma(\delta+1,\pi\delta/2)\\
&+& C^{ab,IR}_k\sum_{(m,n)\not=(0,0)}{2U_2\over \pi|m+nU|^2}
\Gamma\left(1,{\pi
V_2|m+nU|^2\over 2U_2}\right)+\dots\nonumber \ea We have set $\delta=0$ to
all
terms with non-zero momentum. This is justified because we will show that
apart
from the first term, the rest of the sum (in the second term) is finite.
Indeed, the sum over non-zero momenta is finite because it is cutoff by the
incomplete $\Gamma$-function. In fact, for large values of $x$ \be
\Gamma(1,x)=e^{-x}\left[1+{\cal O}\left(1\over x\right)\right] \ee and the
momentum sum is bounded by \be \sum_{(m,n)\not=(0,0)}{2U_2\over\pi|m+nU|^2}
~e^{-{\pi V_2|m+nU|^2\over 2U_2}} \ee which is convergent for $V_2>1$. It
has a
logarithmic divergence $\sim \log V_2$ when $V_2\to 0$ but we always keep
$V_2\geq 1$ in our conventions.
Thus, the pole is given by the first term only
\be
I^{UV}_{k}={4V_2~C^{ab,IR}_k\over \pi\delta}+{\cal O}(\log\delta)\, .
\ee


\appendixB{Appendix B : Calculation of the UV tadpoles for standard
orientifolds}

In this appendix we compute the asymptotic values $C^{UV}_k$ and
$C^{IR}_k$ of $Z_N$ orientifolds.
The relevant $N=1$ sector partition functions are
\be
Z^{99}_{int,k}=Z^{55}_{int,k}=
\prod_{j=1}^3{(2\sin[\pi kv_j])\vartheta\left[
{\alpha\atop \beta+2kv_j}\right]\over\vartheta\left[
{1\atop 1-2kv_j}\right]}\, ,
\ee
\be
Z^{95}_{int,k}=-2(2\sin[\pi k v_1]){\vartheta\left[{\alpha\atop
\beta+2kv_1}\right]\over \vartheta\left[ {1\atop
1-2kv_1}\right]}\prod_{j=2}^3{\vartheta\left[ {\alpha+1\atop
\beta+2kv_j}\right]\over\vartheta\left[ {0\atop 1-2kv_j}\right]}\, ,
\ee
where $k$ runs over $N=1$ sectors, $(v_1,v_2,v_3)$ is the
generating rotation vector of the orbifold satisfying $v_1+v_2+v_3=0$
in order to preserve $N=1$ supersymmetry and the 5-branes are
stretching along the first torus.

Using the property that on $\vartheta$-functions
$i\pi\partial_{\tau}={1\over 4}\partial^2_{v}$, and the Riemmann
identity \be {1\over
2}\sum_{\alpha,\beta=0,1}(-1)^{\alpha+\beta+\alpha\beta}
\vartheta\left[{\alpha\atop
\beta}\right](v)\prod_{i=1}^{3}\vartheta\left[{\alpha+h_i\atop
\beta+g_i}\right] (0)=\vartheta\left[{1\atop
1}\right](v/2)\prod_{i=1}^{3}\vartheta\left[{1-h_i\atop
1-g_i}\right](v/2) \label{rie} \ee in (\ref{fk}), we obtain \be
F_{k}^{99}=F_{k}^{55}={1\over 16\pi^3}\prod_{i=1}^{3}(2\sin[\pi
kv_j])\sum_{i=1}^{3}{\vartheta'\left[ {1\atop
1-2kv_i}\right](0)\over \vartheta\left[ {1\atop
1-2kv_i}\right](0)} \label{form2} \ee \be F_{k}^{59}=-{\sin(\pi
kv_1)\over 4\pi^3}\left[{\vartheta'\left[ {1\atop
1-2kv_1}\right](0)\over \vartheta\left[ {1\atop
1-2kv_1}\right](0)}+{\vartheta'\left[ {0\atop
1-2kv_2}\right](0)\over \vartheta\left[ {0\atop
1-2kv_2}\right](0)}+{\vartheta'\left[ {0\atop
1-2kv_3}\right](0)\over \vartheta\left[ {0\atop
1-2kv_3}\right](0)}\right]\, . \label{form3} \ee Using now \be
{\vartheta'\left[ {1\atop 1-2kv_i}\right](0)\over \vartheta\left[
{1\atop 1-2kv_i}\right](0)}=2\pi\cot(\pi k v_i)+{\cal O}(e^{-\pi
t})\;\;\;,\;\;\;{\vartheta'\left[ {0\atop 1-2kv_i}\right](0)\over
\vartheta\left[ {0\atop 1-2kv_i}\right](0)}={\cal O}(e^{-2\pi
t})\, , \ee we obtain \be C^{99,IR}_k=C^{55,IR}_k={1\over
\pi^2}\prod_{i=1}^{3}(\sin[\pi kv_j]) \sum_{i=1}^{3}\cot(\pi k
v_i)\;\;\;,\;\;\;C^{95,IR}=-{\cos(\pi kv_1) \over 2\pi^2}\, . \ee

For the mass computation, we are interested in the modular
transform of $F_k$. Using
\be
\vartheta'\left[{\alpha\atop
\beta}\right]\left(0,\tau\right)=
-{1\over\tau\sqrt{-i\tau}}e^{i\pi{ab\over 2}}\vartheta'
\left[{\beta\atop -\alpha}\right]\left(0,-{1\over\tau}\right)\, ,
\ee
we can rewrite (\ref{form2}) and (\ref{form3}) as
\be
F_{k}^{99}=F_{k}^{55}=-{1\over 2\pi^3\tau}\prod_{i=1}^{3}(\sin[\pi
kv_j])\sum_{i=1}^{3}{\vartheta'\left[ {1-2kv_i\atop
-1}\right]\left(0,-{1\over \tau}\right)\over \vartheta\left[
{1-2kv_i\atop -1}\right]\left(0,-{1\over \tau}\right)}
\label{form4}
\ee
\be
F_{k}^{95}={\sin(\pi kv_1)\over 4\pi^3\tau}
\left[{\vartheta'\left[ {1-2kv_1\atop -1}\right]\left(0,-{1\over
\tau}\right)\over \vartheta\left[ {1-2kv_1\atop
-1}\right]\left(0,-{1\over \tau}\right)}+{\vartheta'\left[
{1-2kv_1\atop 0}\right]\left(0,-{1\over \tau}\right)\over
\vartheta\left[ {1-2kv_1\atop 0}\right]\left(0,-{1\over
\tau}\right)}+{\vartheta'\left[ {1-2kv_3\atop
0}\right]\left(0,-{1\over \tau}\right)\over \vartheta\left[
{1-2kv_3\atop 0}\right]\left(0,-{1\over \tau}\right)}\right]\, .
\label{form5}
\ee
Defining by $\{kv_i\}$ to be the (positive) fractional part of $kv_i$,
then
\be
{\vartheta'\left[ {1-2kv_i\atop -1}\right]\left(0,-{1\over\tau}
\right)\over \vartheta\left[ {1-2kv_i\atop -1}\right]\left(0,-{1\over
\tau}\right)} =2\pi i\left[\{kv_i\}-{1\over 2}\right]+
{\cal O}\left(e^{-\pi/t}\right)
\ee
and
\be
{\vartheta'\left[ {1-2kv_i\atop 0}\right]\left(0,-{1\over\tau}\right)
\over \vartheta\left[ {1-2kv_i\atop 0}\right]\left(0,-{1\over
\tau}\right)} =2\pi i\left[\{kv_i\}-{1\over 2}\right]
+{\cal O}\left(e^{-\pi/t}\right)\, .
\ee
In the second case, when $\{kv_i\}\in Z$ the limit gives zero.
We must have $|\{kv_i\}-{1\over 2}|<{1\over 2}$. Using now
\be
\eta_k \equiv \sum_{i=1}^3\left[\{kv_i\}-{1\over 2}\right]=
{1\over 2}\prod_{i=1}^{3}{\sin[\pi kv_j] \over |\sin[\pi kv_j]|}\, ,
\ee
we can directly compute (replacing $\tau=it/2$)
\be
C^{99,UV}_k=C^{55,UV}_k=-{1\over \pi^2}
\prod_{i=1}^{3}|\sin[\pi kv_j]|\, ,
\ee
\be
C^{95,UV}_k={\sin(\pi kv_1)\over \pi^2}\eta_k\, .
\ee

\end{document}